\documentclass[12pt, journal, letterpaper]{IEEEtran}
\usepackage{amsmath}
\usepackage{graphicx,subfigure,comment}
\usepackage[table]{xcolor}
\usepackage{url} 
\usepackage{dsfont}
\usepackage{cite}
\usepackage{tabularx}
\usepackage{soul}
\usepackage{todonotes}
\usepackage{balance}

\newcommand{\Fig}[1]{Fig.~\ref{fig:#1}}
\newcommand{\Sec}[1]{Sec.~\ref{sec:#1}}

\begin{document}

\title{AIORA: An AI-Native Multi-Stakeholder Orchestration Architecture for 6G Continuum}
\author{\IEEEauthorblockN{Nuria Molner}\IEEEauthorrefmark{1}\IEEEauthorrefmark{8}, \IEEEauthorblockN{Luis Rosa}\IEEEauthorrefmark{2}, \IEEEauthorblockN{Fulvio Risso}\IEEEauthorrefmark{3}, \IEEEauthorblockN{Konstantinos Samdanis}\IEEEauthorrefmark{4}, \IEEEauthorblockN{David Artuñedo}\IEEEauthorrefmark{5}, \IEEEauthorblockN{Rob Smets}\IEEEauthorrefmark{6}, \IEEEauthorblockN{Tarik Taleb}\IEEEauthorrefmark{7}, \IEEEauthorblockN{David Gomez-Barquero}\IEEEauthorrefmark{1}\\
\IEEEauthorblockA{\IEEEauthorrefmark{1} iTEAM Research Institute - Universitat Polit\`ecnica de Val\`encia, Spain} \IEEEauthorblockA{\IEEEauthorrefmark{2} One Source Consultoria Informatica LDA, Portugal} \IEEEauthorblockA{\IEEEauthorrefmark{3} Politecnico di Torino, Italy} \IEEEauthorblockA{\IEEEauthorrefmark{4} Lenovo Deutschland GMBH, Germany} \IEEEauthorblockA{\IEEEauthorrefmark{5} Telefónica Innovacion Digital SL, Spain} \IEEEauthorblockA{\IEEEauthorrefmark{6} Nederlandse Organisatie voor Toegepast Natuurwetenschappelijk Onderzoek TNO, Netherlands} \IEEEauthorblockA{\IEEEauthorrefmark{7} Ruhr University Bochum, Germany \vspace*{-1cm}}
\thanks{\IEEEauthorrefmark{8}corresponding author: numolsiu@iteam.upv.es}
}
\maketitle

\begin{abstract}
This paper elaborates on a novel AI-native architecture for emerging 6G systems harnessing open APIs, along with supporting mechanisms to empower intelligent and coordinated orchestration of edge-cloud continuum resources. The AIORA architecture facilitates a seamless creation, life-cycle management, and exposure of services in multi-segment heterogeneous environments. It integrates new breeds of tools and advanced technologies to enable zero-touch management of an edge-cloud continuum, building on top of the 3GPP Edge Enablement Layer and the respective connectivity models, allowing to cater to the high flexibility, availability, efficiency, reliability, and resilience needs of the future 6G services and applications. Several ongoing industry initiatives—such as ETSI MEC for edge computing platforms, the GSMA Operator Platform for multi-operator service federation, and CAMARA for cross-operator API standardization—demonstrate the growing momentum towards integrated frameworks where edge, cloud, and network resources can be seamlessly orchestrated. Our proposed AIORA architecture not only aligns with these initiatives but also extends them by leveraging a multi-segment virtual continuum concept and nested AI-driven closed loops for real-time optimization.
\end{abstract}


\section{Introduction}
\label{sec:intro}

The European Vision~\cite{6Geurope} 
for the 6G Network Ecosystem anticipates a landscape built on three foundational pillars: immersive communication, cognition, and digital twinning. These pillars embrace the concept of 6G connected intelligence, ubiquitous computing and connectivity, as elaborated in IMT-2030 \cite{ITU2023}. Thus, the realisation of upcoming 6G use cases \cite{HexaX} requires a sustainable, seamlessly interconnected infrastructure across multiple system segments across heterogeneous technologies and/or administrative domains and stakeholders.

Emerging 6G network architectures should be also envisioned to enable interactive and distributed services closer to the user at the network edge. To accomplish this, certain enhancements are introduced to extend the Service Based Architecture (SBA) towards the radio, introducing the notion of Radio Access Network (RAN) and core convergence. RAN-core convergence brings selected features of the core network at the edge, allowing a customized merge of core with radio functionality to enable a flexible 6G architecture in where radio services can directly be consumed by the core network and vice versa \cite{6G-Cloud}.

In turn, this calls for a novel framework capable of supporting secure and fully automated service orchestration and multi-objective optimisation on top of a ubiquitous (distributed~\cite{6Groadmap, EUinfrastructures}) edge-to-cloud virtual continuum. Such virtual continuum can be anticipated as an interconnected array of virtual resources, which can either (1) span over a singular segment – e.g., Kubernetes cluster  –  or (2) be distributed across multiple segments, in which case, the virtual continuum shall also include the essential components that facilitate seamless connectivity across virtual resources in different segments composing such environment, ensuring cohesive operation and resource utilization.

Currently, there is a lack of a holistic approach to solve this problem end-to-end (E2E), as it is composed of multiple smaller and yet complex sub-problems with different actors involved that are being addressed independently. On the one hand, there is a layered approach for software networks and edge locations that do not scale when multiple stakeholders are involved. Layers include edge applications and arrangements, containers, hypervisors and hardware solutions. For each layer, there are (partial) solutions for a single stakeholder, as each stakeholder has complete control of its resources. However, dynamically allocating resources becomes more complex as new player becomes part of the distributed network deployment. Business relationships affect privacy, data location, and resource availability, among others, and, thus, create barriers to transparently sharing resources, ultimately impacting the possibility of having a ubiquitous orchestration.

In parallel, the ETSI MEC (Multi-access Edge Computing) framework provides a standardized environment for running edge applications close to the end user, offering service APIs (Application Programming Interface) (e.g., radio network information, location) to facilitate context-aware orchestration. Meanwhile, the GSMA Operator Platform advocates a global model for edge service federation, enabling consistent cross-operator offerings and unified access to edge resources. Although these efforts significantly advance edge-cloud integration, they often focus on single-stakeholder scenarios or require complex, bilateral federation agreements that do not entirely address dynamic, AI-driven orchestration across multiple administrative domains.  This exposes the need for interfaces communicating and orchestrating multi-stakeholder scenarios.

On the other hand, several pieces need to be orchestrated seamlessly, integrating different solutions: the management of Edge Computing resources defined by Telco Operator Platform, the APIs for managing Edge Server resources and Applications defined by the 3GPP EDGEAPP Framework, Kubernetes Multi-cluster topologies, User Plane Function (UPF) topologies for accessing Edge services, Network status information, Customers’ Location and Requirements defined by 3GPP Connectivity models, etc. The complexity required to combine business relationships, resource availability, network status, and customer requirements in real-time can only be achieved by a novel artificial intelligence (AI) framework that covers all dimensions natively.

To this end, this paper defines a unified 6G AI-native architecture, AIORA, contributing to identifying and addressing the relevant challenges and current implementations and solutions to be extended for an architecture that works hand in hand with the current and emerging European edge infrastructure to handle appropriately its principles of relevance for the European society~\cite{6Groadmap}.

\Sec{sota} describes state of the art of E2E 6G orchestration problem,~\Sec{architecture} elaborates the AIORA AI-native multi-stakeholder 6G architecture.~\Sec{scenarios} describes the business scenarios considering 3GPP EDGEAPP Framework.~\Sec{arch_capabilities} provides an analysis on the capabilities related to the proposed AIORA architecture elaborating how to assist the emerging 6G services and applications. Finally,~\Sec{conclusion} concludes the paper.      

\section{State of the Art}
\label{sec:sota}

The AIORA architecture is designed to address the E2E 6G orchestration problem, which is composed by multiple smaller yet complex sub-problems with different actors involved, for some of which independent solutions exist.

To satisfy the requirements of 6G, there is a growing need to establish simple, dynamic, automated, and facilitated ways to expose network management services and APIs across operators. In 6G, one common network interface should exist to integrate non-public networks (NPN) management with Operational Technology (OT) networks. Current standards and solutions, such as 3GPP Common API Framework (CAPIF~\cite{capif}), 3GPP Service Enabler Architecture Layer for Verticals (SEAL~\cite{SEAL}), CAMARA~\cite{camara}, target the publishing, discovery and usage of APIs which could in principle enable more flexible and dynamic management and monitoring of the network and services to third parties, provide improved control, auditability and ease the management of the exposed APIs to operators. Concerning edge enablement, SA6 (Service and System Aspects) has created the Edge Computing Application Framework (EDGEAPP TS 23.558), integrated with Common API Framework (CAPIF~\cite{capif}).
This framework follows these four Architecture Principles: (\textit{i}) Application Portability: Minimise changes to the Applications in UEs and Application servers; (\textit{ii}) Service Differentiation: capability to enable/disable Edge Computing features; (\textit{iii}) Flexible Deployment: Mobile Network Operators (MNOs) can support multiple Edge Computing Providers; (\textit{iv}) Interworking with 3GPP Networks.

Regarding the management and orchestration of resources, Liqo-based Kubernetes orchestration~\cite{liqo} provides the basic building blocks to handle the lifecycle of the virtual computing continuum, allowing services running in different clusters to interact and communicate as if they were in the same physical infrastructure. ETSI ZSM (Zero-touch network and Service Management) targets the zero-touch management of resources and provides guidelines for standardized usage with the concept of closed loops for service management~\cite{ZTM}. ETSI ENI (Experiential Networked Intelligence)~\cite{ENI} provides a reference and insights into AI and machine learning integration in network management. Additionally, SYLVA~\cite{sylva} Linux Foundation is defining an integrated Software Stack for Telco Edge as a fundamental step to Telco Cloud and Edge homogenization and sustainability. 

Some of the existing solutions are open-source, but others are proprietary, and many of these solutions for independent pieces are incompatible. Thus, there is an urgent need for a holistic approach to solve the problem E2E, and the first step is a unified architecture that seamlessly holds the different pieces.

\section{AIORA AI-Native Multi-stakeholder 6G Architecture}
\label{sec:architecture}
Novel real-time and interactive 6G applications need to migrate easily across different edge clouds with assured network connectivity in terms of peak throughput, reliability, and latency while preserving users’ data privacy and data location control.

\begin{figure*}[t]
    \centering
    \includegraphics[width=.95\textwidth]{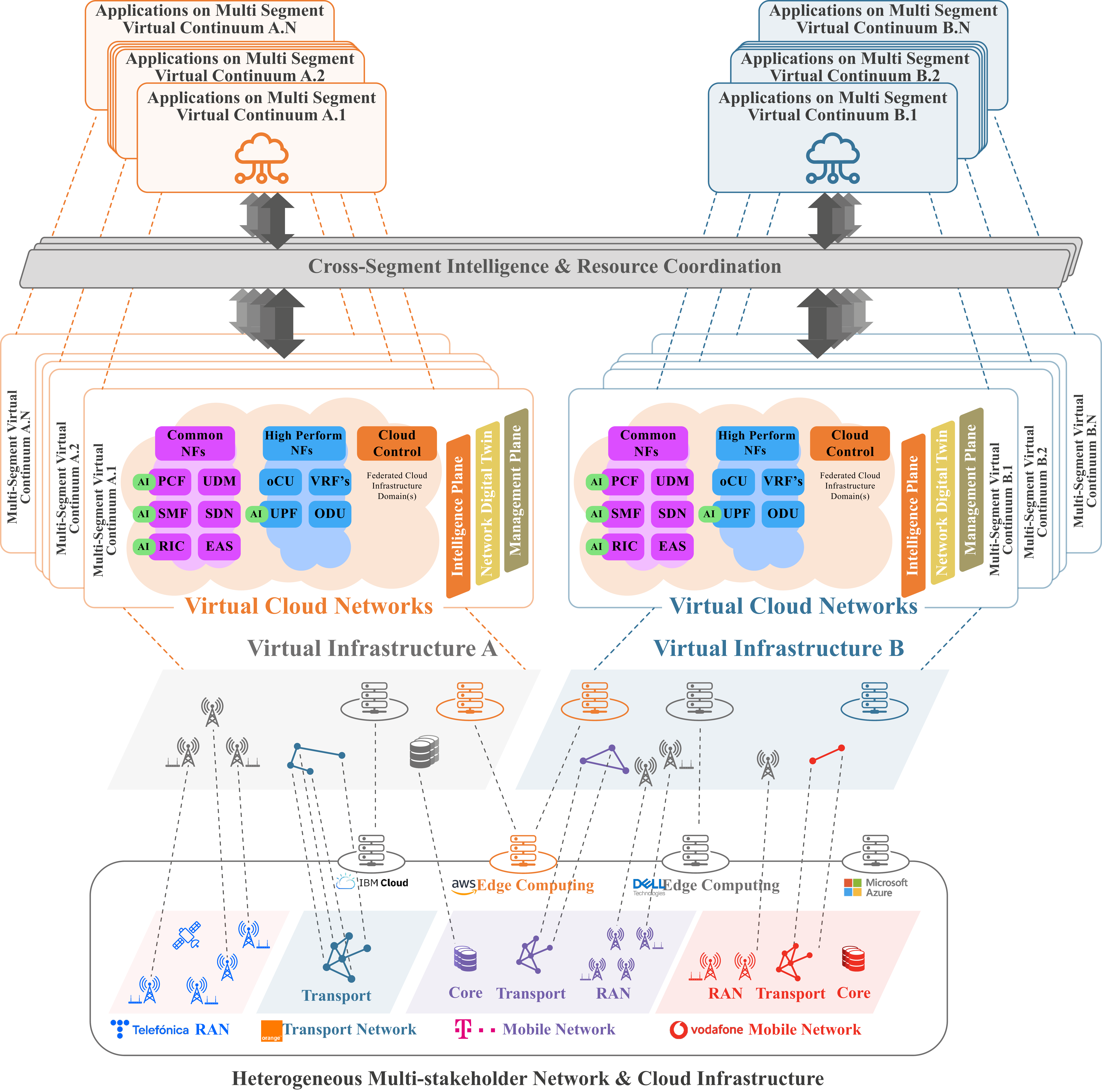}
    \caption{The envisioned AIORA architecture showing the composition of multi-segment virtual continuums based on multi-stakeholder heterogeneous network and cloud infrastructure resources, with each multi-segment virtual continuum being configured with a customised user, control, cloud, management, and intelligence planes. A cross-segment intelligence and resource coordination substrate resolves conflicts among multi-segment virtual continuums, assuring a service targets harmonisation related to the applications enabled on the top.
        \label{fig:architecture}
    }
\end{figure*}

AIORA architecture illustrated in~\Fig{architecture} addresses scalability, performance, sustainability, and operational issues that will arise from the need to assemble applications, i.e., compose applications with different components stretched across federated multi-segment edge environments. In addition, it integrates principles from ETSI MEC, where a platform manager can manage edge application lifecycles and service APIs, and leverages the GSMA Operator Platform approach to federate resources across multiple operators. Hereinafter, we refer to CAMARA-style APIs as standardized, network-aware APIs developed through the CAMARA initiative that expose network capabilities (like quality of service, location, or slicing) to application developers in a uniform, operator-agnostic way, enabling seamless integration of telecom features into digital services across different networks and vendors. Thus, by adopting CAMARA-style APIs aligned with 3GPP CAPIF and SEAL frameworks, the solution enables a uniform developer interface for dynamic service deployment and closed-loop automation, while still preserving the distributed (and de-centralized) nature of telco services.

It is worth mentioning that the AIORA architecture is designed to complement --not conflict with-- existing and emerging standards such as 3GPP SA6’s SEAL framework and CAPIF APIs, Open-RAN (O-RAN)’s RIC architecture, and the International Mobile Telecommunications (IMT)-2030 vision for network softwarization. Rather than duplicating or replacing these frameworks, our approach integrates them by embedding AI-driven orchestration, automation, and intelligence across layers. It leverages CAPIF for service exposure and discovery, enhances SEAL’s service enablement with AI-native optimization, and interfaces with O-RAN’s RIC to support adaptive, edge-intelligent operations through higher-level AI agents. Aligned with IMT-2030’s goals, the AIORA architecture builds on principles of softwarization, intent-based management, and cloud-native design, forming an intelligence fabric that unifies and extends existing capabilities across the 6G ecosystem.

Through these enhancements, we make such applications mobile, to interwork and jointly materialise on the user and business requirements. Scalability issues manifest themselves in (\textit{a}) unbridled growth of relations between network functions in the domains of the involved MNOs, and applications providers, and (\textit{b}) the complexity of interaction, i.e., the number of messages that must be exchanged that span numerous network functions, among the involved business partners. Performance issues relate to fulfilling the required Quality of Service (QoS) needed by the application(s) and Quality of Experience (QoE) expected by the end user, allowing the applications to be interconnected over network and computing resources with low latency and ample bandwidth. Sustainability aspects such as energy consumption and carbon footprint requirements must be addressed. The AIORA architecture enables reduced energy consumption and cost through fine-grain control of cloud resources needed by network functions and application servers, and the capability to optimize service placement (with energy/carbon-related constraints) over a wide infrastructure. Finally, this architecture offers operational advantages, supporting the legacy 5G infrastructure and 6G additions to come and ensuring that 6G capabilities lead to operational excellence in monitoring, fault isolation, security, administration, and management, and be compatible with organisational processes.

The proposed AIORA architecture addresses the control, relocation and scaling of network management facilities, application assembly (i.e., how applications components are interrelated to composed complex applications), network functions and application servers in and across federated mobile, private and application (cloud) segments. This can assist the assurance of end-user QoE when it becomes beleaguered by events, such as but not limited to end-user mobility, server maintenance shutdown, high network loads, and cyber-attack events.

Cross-segment intervening measures and continuous optimisation must ensure the QoE of the end users. This requires a new AI-native control and management framework. The notion of AI shall be embedded on the one hand in the control and management functions, including core network and radio functions. On the other hand, AI-nativeness shall introduce decision and execution intelligence to address flexibility by continuously considering the different platforms, environments, applications, and end-user behaviour. In other words, AI shall enhance protocol mechanisms and decision-making while simultaneously enabling control closed loops to be created and deployed efficiently, i.e., without conflicts in a cooperative environment, while allowing for nested closed-loop control operations.

Multi-segment virtual continuums must be tailored to the requirements of MNOs, edge computing service providers and application service providers. Each multi-segment virtual continuum may support different applications with similar service characteristics, e.g., geographical area, mobility or QoS parameters. Modifications made to accommodate the needs of a certain multi-segment virtual continuum may impact others that use the same network or cloud resources, especially when modifications occur simultaneously across more than one multi-segment virtual continuum. Cross-segment intelligence and resource coordination is a logical function introduced to resolve conflicts and take decisions based on the virtualised resource information related to the underlying infrastructures and may trigger an interaction with the application providers for negotiating the service quality targets to provide effective coordination and harmonisation.

This system is designed to ensure interoperability across proprietary and open-source orchestration components in multi-stakeholder domains, our AIORA architecture adopts a modular, standards-aligned, and API-centric design. We rely on open, standardized interfaces -- aligned with the specifications of 3GPP, ETSI Network Function Virtualization (NFV), O-RAN, and the Linux Foundation -- to facilitate plug-and-play integration and reduce the risk of vendor lock-in. By employing intent-based orchestration and AI-driven mediation layers, the AIORA architecture can dynamically interpret and translate policies or configurations across heterogeneous systems. Furthermore, the inclusion of interoperability gateways, model-driven architectures (e.g., YANG, TOSCA), and open telemetry standards (e.g., OpenTelemetry, Prometheus) allows both proprietary and open-source components to communicate effectively. Through these design principles, we embed real-world challenges of fragmentation, enabling seamless, policy-compliant orchestration across diverse stakeholders while fostering innovation and openness.

The AI-Native implementation of the AIORA architecture builds on top of the multi-segment managed infrastructure and its Digital Twin, and it relies on an AI-Native containerised pipeline where AI \& Generative AI Models live and are trained using data prepared by real-time dataset generators. This training data is obtained from monitoring, analytical, orchestration, and actuation functions and will be used to train both AI and Generative AI models. The AI models will enhance monitoring, analytics, orchestration, and actuation operations to provide classification and prediction capabilities to the existing management functions. Meanwhile, Generative AI models will be used to generate the optimal configurations that will enrich monitoring, analytics, orchestration, and actuation sections for continuous optimal enhancement. The AI and Gen AI models, together with infrastructural elements and their digital twin counterparts, form one or more closed loops. The closed-loop coordinator is responsible for initialising and supervising the closed loops to guarantee stable, intent-driven, and, preferably, optimised operation, even when closed loops are nested or intertwined. In this sense, the coordination action shall consider the actuation decision, the managed entities, and the relationship between the closed loops to select the actuation decision to be enacted. 

The AI-native closed-loop control system leverages a combination of reinforcement learning, transformers, and GNNs for analytics, with hierarchical orchestration and multi-objective optimization for decision-making. Actuation is handled via standardized APIs from 3GPP, ETSI, and O-RAN frameworks. The AI pipeline supports real-time data ingestion, federated and centralized training, and explainable model deployment through MLOps tools. At scale, challenges like training overhead and latency are addressed through model compression, edge inference, and hybrid pre-trained/continual learning strategies. The AIORA architecture is designed to balance intelligence and efficiency, enabling scalable, real-time automation across heterogeneous domains.

The AIORA architecture addresses security by using the security-by-design strategy, which relies on a threat analysis of the different architectural alternatives and their operation. Different layers of security services can be integrated to enable confidentiality, integrity, and availability of services. Such layers include different threat security functionalities, as cryptography primitives (classical, post-quantum and hybrid), and AI-based closed loops ensuring different security levels that can be mapped on the different segments to fulfil the security application needs.

\section{6G Multi-Stakeholder Business Scenarios: A 3GPP EDGEAPP Approach}
\label{sec:scenarios}

\begin{figure*}[t]
    \centering
    \includegraphics[width=\textwidth]{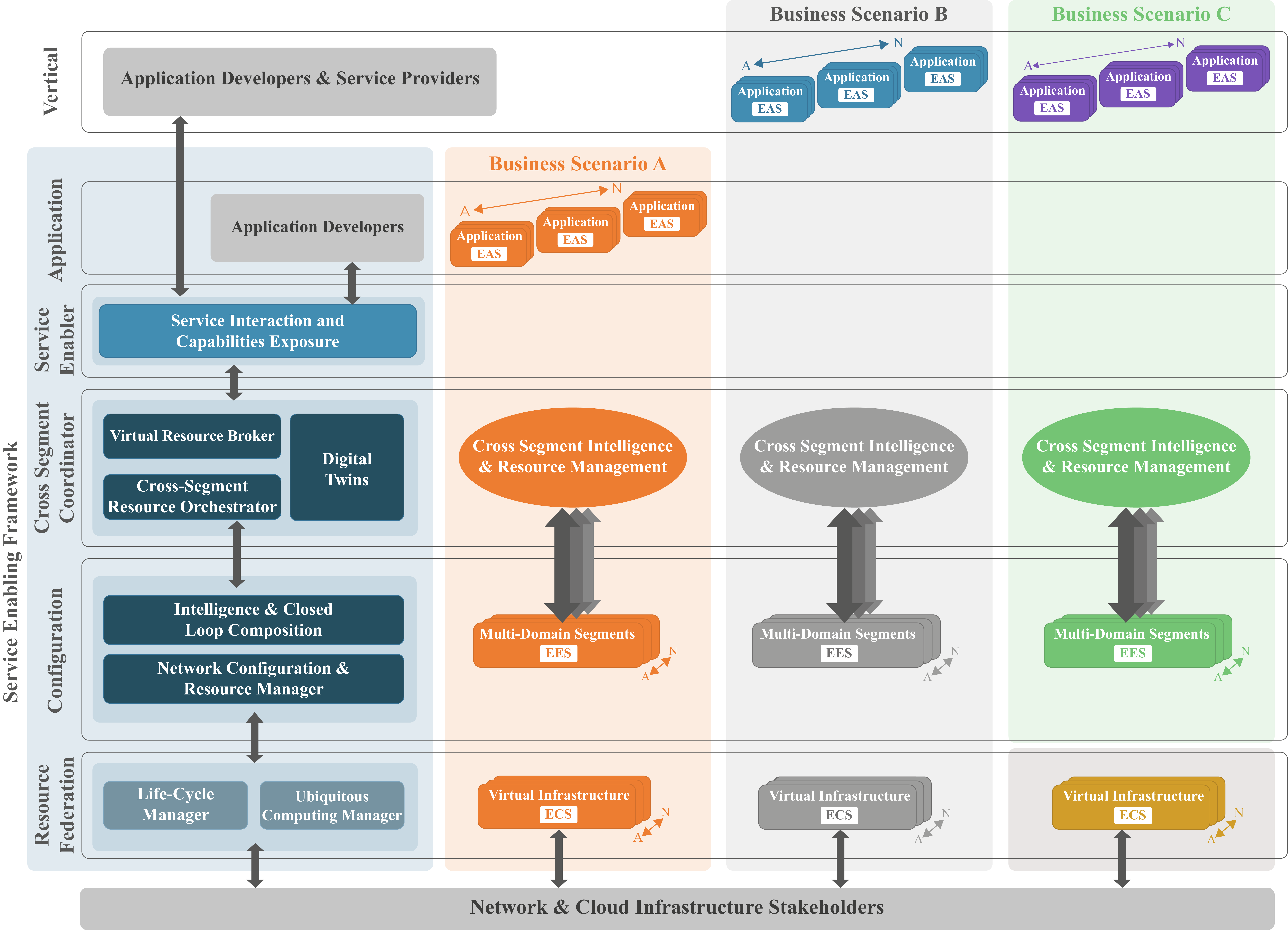}
    \caption{Service Enabling Framework that offers the tools for facilitating services across a heterogeneous multi-stakeholder cloud and network infrastructure and shows the relation with the 3GPP SA6 WG business scenarios: (i) Business Scenario A: Single business operator that offers and manages all entities of the system, (ii) Business Scenario B: An application provider offers the application servers and an MNO or an edge computing service provider offers the virtual infrastructure and the multi-segment virtual continuums, while a single business operator manages all entities of the system, and (iii) Business Scenario C: An application provider offers the application server and an MNO, or an edge computing service provider offer the virtual infrastructure and the multi-segment virtual continuums, while an MNO is responsible for the management of the virtual infrastructure and service provider for the virtual network on the top including the applications.
        \label{fig:architecture-business}
    }
\end{figure*}

The proposed AIORA architecture aims to support different edge data networks following the paradigm of 3GPP SA6 WG, as illustrated in \Fig{architecture-business}, by enabling the deployment of:
\begin{itemize}
    \item Edge Application Servers (EAS) resident in the edge data network vertical segment or application side inside the multi-segment virtual continuum business provider.
    \item Edge Enablement Server (EES) in the multi-segment virtual continuum to assist and configure the edge, and support EAS by offering instantiation, interface invoker and network exposure.
    \item Edge Configuration Server (ECS) in the virtual infrastructure for provisioning edge configuration to the client, support registration information for EES, and for providing a network interface.
\end{itemize}
Notably, integrating an ETSI MEC Platform Manager within each multi-segment virtual continuum can further streamline resource orchestration and service discovery, while the GSMA Operator Platform can federate multiple MEC-enabled domains. In scenarios where operators or service providers must expose capabilities to developers, a CAMARA-based API layer ensures a uniform approach to discovering and invoking EES, ECS, and related features.

Three different business scenarios, considering 3GPP SA6 TS 23.558
, are supported by the AIORA architecture as shown in \Fig{architecture-business} (for each scenario, the colour of the components indicates the business entities that offer them and the background colour of the business operator that provides the management).
They are as follows:

\begin{itemize}
    \item Business scenario A: A single business operator, which can be either an MNO or an application service provider, offers and manages the virtual infrastructure (i.e., the multi-segment virtual continuums), and the application server on top of it; 
    \item Business scenario B: A business operator, which can be an MNO or an edge computing service provider, offers the virtual infrastructure (i.e., the multi-segment virtual continuums), while an application service provider offers the application server. The entire system is then managed by the former MNO or the edge computing service provider, respectively; 
    \item Business scenario C: An MNO offers and manages the virtual infrastructure. At the top, an edge computing service provider offers multi-segment virtual continuums, while an application service provider offers the application server. The former edge computing service provider manages the multi-segment virtual continuums and the respective application server.
\end{itemize}

The AIORA architecture consists of multiple segments creating virtual continuums that leverage the appropriate subset of existing infrastructure-level resources. It assumes that an intermediate multi-segment virtual continuum business provider, i.e., an MNO or a service provider, collects physical resources from other MNOs and edge computing service providers based on service level agreements.

The life-cycle management of multi-segment virtual continuums is a complex issue addressed by introducing the service-enabling framework, as depicted in \Fig{architecture-business}. It offers a set of mechanisms and tools to enable application developers or service providers to deploy applications that can run in any location of the virtual continuum, with the capability to migrate following the user needs and the dynamics of the underlying infrastructure across heterogeneous resources. Leveraging the benefits of AI/ML, the service-enabling framework introduces automation by employing nested control closed loops. The service-enabling framework aggregates the physical resources and offers a unified view of the virtualised federated cloud and network environment using the following functions:
\begin{itemize}
    \item Life-cycle manager takes care of the preparation, instantiation, activation, maintenance, modification, and termination of resources.
    \item Ubiquitous computing manager that ensures the interoperation and management of the cloud resources and corresponding network resources. 
    \item Infrastructure configuration and resource manager is responsible for providing the configuration of federated radio access, cloud, control plane and user plane, i.e., in terms of the radio base stations coverage, composition of network functions, cloud services (e.g., EAS discovery, network capability exposure using CAPIF, service continuity) and control mechanisms. In addition, it configures the management and orchestration tools needed in the management plane and assembles the closed loops and algorithms for enabling the intelligence plane.
    \item Intelligence and closed loop composition deals with creating controlled closed loops in a federated resource environment, i.e., selecting the monitoring that shall be used, the analytics to provide the insights, the decision entities, and the orchestrator – actuator that would enforce decisions. It also offers AI/ML algorithms and the respective data preparation and training tools and provisions the desired native AI capabilities to NFs and application services. 
\end{itemize}

Multi-segment virtual continuums may compete for the same resources or employ intelligent and management mechanisms that impact each other. To resolve these issues, cross-segments intelligence and resource coordination are introduced by adopting the following tools:

\begin{itemize}
    \item Multi-segment resource orchestrator resolves resource conflicts and other network management-related issues.
    \item Virtual resource broker collects and maintains federated physical resources (i.e., network, computing, storage) to facilitate service feasibility, check and other related resource operations.
    \item Network Digital Twin provides a view of how the network resources, services, and applications perform. It is built using information from the digital twins exposed by external segments and information obtained from the local segment, i.e., from telemetry and performance monitoring data.
\end{itemize}

Multi-segment virtual continuums are composed by considering the service requirements introduced by the respective applications. The service enabler layer allows the deployment of applications related to specific multi-segment virtual continuums service interaction and capabilities exposure, which is responsible for enabling the composition of a service by analysing the service requirements and checking if the desired service can be supported considering the virtual infrastructure resource availability by interacting with the virtual resource broker. In addition, it enables a set of tools to allow the application and service provider to interact with the multi-segment virtual continuum business provider. The service-enabling framework executes various workflows designed to achieve a desired response from its own technology and external segments. These workflows can be triggered periodically or upon an event, i.e., on-demand, to perform various optimisations.


\section{AIORA AI-Native Multi-Stakeholder 6G Architecture: A Capability Analysis}
\label{sec:arch_capabilities}

This novel AIORA architecture goes beyond the current 5G architecture, which typically relies on a network slice-level federation of the resources with limited orchestration capabilities. Network slicing allows to create multiple virtualized and isolated networks --called slices-- on a shared physical infrastructure. Each slice is tailored to specific use cases, with predefined performance characteristics. 5G slicing is largely static or semi-dynamic, centrally orchestrated, and constrained by pre-configured templates~\cite{NetworkSlicing}. By embedding ETSI MEC’s edge orchestration mechanisms and leveraging the GSMA Operator Platform for cross-operator resource sharing, we enable far more granular and dynamic resource management. Moreover, adopting CAMARA-based APIs ensures that developers can consistently access and manage edge functions across heterogeneous domains, while AI-driven closed-loop control provides automated real-time optimizations. Consequently, we enable intelligent, pervasive, fine-grained resource sharing, along with federated access and exposure to real-time monitoring insights.

Building over 5G architectures~\cite{5G-6G-SDN} based on Software Defined Networking (SDN) and application scenarios with mobility management approaches in mobile networks, the AIORA architecture exploits the limitations of the previous generation. The cornerstone behind this novel architecture is the leverage of virtual resources and service groupings on top of the virtualised infrastructure, referred to as multi-segment virtual continuums. These multi-segment virtual continuums enable the creation of sets of network, computing and storage resources, application and services (including network functions), and policies tailored for a specific set of application requirements in terms of network capacity, coverage, latency, etc. In this vision, each multi-segment virtual continuum shall contain its own: (\textit{i}) Radio access, composed of a set of base stations and radio units; (\textit{ii}) 6G core network control plane and multipurpose user plane, which could be customised as per the features of the target application(s); (\textit{iii}) Cloud and (near/far) edge computing resources where the different application functions will be deployed; (\textit{iv}) Intelligence and management planes implementing the per-virtual continuum Native AI control-loops for radio, network and compute resource management and optimisation; (\textit{v}) Native AI tools for algorithm lifecycle management (i.e., training and execution); and the virtual continuum-specific management tools enforcing the lifecycle management and optimisation actions.

To maximize efficiency and interoperability, the creation, management, and coordination of multi-segment virtual continuums are orchestrated by a dedicated multi-segment virtual-continuum business provider, which: (\textit{i}) Provides a high-level and aggregated view of the virtual continuum towards application developers and service providers; (\textit{ii}) Offers high-level application lifecycle-management primitives; (\textit{iii}) Offers automated management of the heterogeneous virtualised-infrastructure segments; and (\textit{iv}) Offering automated coordination and conflict resolutions of closed control loops across multi-segment virtual continuums.

The AIORA architecture represents a transformative evolution from the reactive, policy-driven nature of 5G to a proactive, intent-based framework powered by distributed intelligence. By embedding AI agents across the RAN, core, and edge, it enables closed-loop automation, real-time analytics, and continual learning, which together unlock dynamic service orchestration and fine-grained optimization. This shift allows the network to scale more efficiently across highly diverse use cases and stakeholders, overcoming the centralized bottlenecks typical of 5G. Energy consumption is intelligently managed through predictive workload placement and adaptive control strategies, leading to significant efficiency gains, especially in IoT-heavy or edge-intensive environments. Meanwhile, latency is reduced by enabling localized, AI-driven decision-making closer to the user, supporting emerging ultra-low-latency applications such as immersive Extended Reality (XR) and tactile communication. Overall, AI-native 6G delivers a more responsive, scalable, and energy-aware infrastructure compared to traditional 5G systems.

The AIORA architecture not only advances technological innovation but also drives economic sustainability. By optimizing resource utilization and energy consumption, it significantly reduces both deployment and operational costs. Furthermore, it fosters an open, interoperable ecosystem that encourages collaboration, accelerates market growth, and lowers entry barriers for new players (newly offered cloud infrastructure can now be absorbed as a resource and become part a virtual continuum). Its inherent flexibility ensures seamless adaptation to emerging technologies, protecting long-term investments and paving the way for future innovation. Ultimately, this approach stimulates technological progress, fosters new value chains, and unlocks novel business models, catalyzing innovation and safeguarding long-term investments.

\section{Conclusion}
\label{sec:conclusion}
This paper presents AIORA, a novel AI-native, distributed, multi-segment, and multi-stakeholder architecture tailored for 6G applications, which are expected to support real-time interactivity, with the requirement to migrate easily across different edge clouds with assured network connectivity quality in terms of peak-throughput, reliability, and latency while preserving users-data privacy and data location control. 

The AIORA architecture is centered on the usage of groupings of virtual cloud and network infrastructure resources and services of multiple system segments (e.g., public and private, terrestrial and satellite, and inter-operation among MNOs, service and application providers) on top of a virtualised infrastructure, referred to as multi-segment virtual continuums. It enables a more pervasive and fine-grained sharing of the resources of the current 5G architecture, federated at the network slice level, and federated access and exposure of the monitoring information.

In summary, our AI-native orchestration framework — by embracing ETSI MEC for edge service enablement, the GSMA Operator Platform for multi-operator federation, and CAMARA APIs for standardized network capability exposure — can expedite the industry’s move toward open, interoperable, and continuously optimized 6G ecosystems. The proposed multi-segment virtual continuum concept and its supporting intelligence layer provide a unified, future-proof foundation capable of accommodating emerging technologies and diverse stakeholder requirements.

\section*{Acknowledgment}
This work has been partially supported by the Spanish Ministry of Economic Affairs and Digital Transformation and the European Union-NextGenerationEU through the UNICO 5G I+D ADVANCING-5G-TWINS (Grant No. TSI-063000-2021-112, TSI-063000-2021-113, TSI-063000-2021-114) projects; and by the European Union’s Horizon Europe programme for Research and Innovation through the 6G-SANDBOX project (Grant No. 101096328) and the 6G-Path project (Grant No. 101139172).
The paper reflects only the authors’ views. The European Commission and the Spanish Ministry are not responsible for any use that may be made of the information it contains.

\balance
\bibliographystyle{IEEEtran}
\bibliography{refs}

\end{document}